\documentclass[12pt]{article}
\usepackage{epsfig}
\begin{document}
\begin{center}
{\Large {\bf Evidence for the reality of singular configurations in
$SU(2)$ gauge theory.}}

{
\vspace{1cm}
{B.L.G.~Bakker$^a$, A.I.~Veselov$^b$, M.A.~Zubkov$^b$  }\\
\vspace{.5cm}
{ \it
$^a$ Department of Physics and Astronomy, Vrije Universiteit, Amsterdam,
The Netherlands \\
$^b$ ITEP, B.Cheremushkinskaya 25, Moscow, 117259, Russia
}}
\end{center}
\begin{abstract}
We consider the $SU(2)$  lattice gauge model and investigate
numerically the continuum limit of the simple center vortices which are
singular configurations of the gauge fields. We found that the
vortices remain alive in the continuum theory.  Also we investigate the
Creutz ratio and found that for all $\beta$ it vanishes for those field
configurations which do not contain the simple center vortices inside
the considered Wilson loop. It leads us to the conclusion that these
singular field configurations play a real role in the continuum
theory.

\end{abstract}


\newpage

\section{Introduction}

In the perturbative analysis of field theory there is no
question about whether singular field configuration play some role
in physics or not.  The phenomenological rules for the calculation of
gaussian functional integrals give us the expressions for the
perturbative expansion of the Green functions. The functional
integral itself as a mathematical concept is defined as the integral
over the Haar measure on some functional space.  However, we do not know
what is that functional space for quantum field theory.  We even do
not know what is the space, the integral over which gives us the
correct perturbation expansion for the case of Gaussian integrals. (It
means: we do not know what is the space, for which the analogue of the
expression for the Gaussian integral $\int dx e^{-x^2 + 2iy} = e^{-y^2}$
is valid. For a review of progress made in recent years see, e.g.,
Ref.~\cite{Streit}.)

The universal way to define the functional integral is lattice theory.
In lattice theory there is no question:  ``What is the functional
space, $C_{\infty}, C_1, C$ or other?'' We consider the point of the
second order phase transition and propose to use this point as the
window from the lattice to the continuum. Thus the continuum functional
space is defined in a very simple way. It contains only such
configurations, which survive when we are jumping through this window
from the lattice to the continuum.

Contrary to the situation in perturbation theory, in
nonperturbative field theory the question: ``What kind of field
configurations survive?'' is very sensible, because the topological
properties of the vacuum strongly depend upon the functional space.
For example, if singular configurations are forbidden, there are
no monopoles in pure gauge theories without abelian projection.  But
the existence of such topological objects changes essentially the
phenomenology of the theory.

There have been many attempts to understand what kind of singular
fields play a role in the continuum theory (see for example,
\cite{C,D,F}). Now we add one more work to this list. We prove that
the singular simple center vortices survive in the continuum limit and
play an important role in the confinement picture.

A few facts about the Simple Center Projection:  It was proposed to
illustrate the connection between topological string-like excitations
and the confinement mechanism \cite{BVZ00}.  (For a recent review of
other center projections see Ref.~\cite{Greensite}.) Numerically this
procedure is much more simple than the Maximal Center projection
considered before.  Moreover this procedure is gauge invariant. It was
noticed that the simple center vortices carry singular field strength
in the continuum limit.

Before we carried out the present investigation, we supposed that one
of the following three possibilities may take place:

1. The vortices disappear in the continuum limit;

2. The vortices remain in the continuum limit but do not influence the
physical results;

3. The vortices survive the continuum and play a real role in the
dynamics.

In the present work we found that the third possibility is realized.
It means that the physical functional space should contain
singular fields of such a kind.

\section{The Simple Center Vortex}

Let us recall the definition of the Simple Center Projection.
We consider $SU(2)$ gluodynamics with  the Wilson action
\begin{equation}
 S(U) = \beta
\sum_{\mathrm{plaq}} (1-1/2 \, \mathrm{Tr} U_{\mathrm{plaq}}).
 \label{eq.011}
\end{equation}
The sum runs over all the plaquettes of the lattice. The plaquette
action $U_{\rm plaq}$ is defined in the standard way.

We consider the plaquette variable
\begin{eqnarray}
 z_{\rm plaq} = 1 , & {\rm if} &  {\rm Tr} \, U_{\rm plaq} < 0, \nonumber\\
 z_{\rm plaq} = 0 , & {\rm if} &  {\rm Tr} \, U_{\rm plaq} > 0.
 \label{eq.020}
\end{eqnarray}

We can represent $z$ as the sum of a closed form $d N$ \footnote{We use
the formulation of differential forms on the lattice, as described for
instance in Ref.~\cite{forms}.} for $N\in\{0,1\}$ and the form $2m +q$.
Here $N = N_{\rm link}$, $q\in \{0,1\}$, and $m\in {\sf Z} \!\! {\sf Z}$.

\begin{equation}
 z = d N + 2m +q .
 \label{eq.030}
\end{equation}
The physical variables depending  upon $z$ could be expressed through
\begin{equation}
 {\rm sign} \, {\rm Tr} \, U_{\rm plaq} = \cos (\pi (dN +q)) .
 \label{eq.040}
\end{equation}
$N_{\rm link}$ is the center projected link variable.

For each $U_{\rm link}$ and $N_{\rm link}$, $q$ is defined as a $Z_2$
function of $N$ and $U$
\begin{equation}
 q = q(N,U) .
 \label{eq.050}
\end{equation}
For each $U$ we minimize $\sum_{\rm link} q$ with respect to $N$, which
is fixed locally. All links are treated in this way and the procedure
is iterated until a global minimum is found.

Geometrically this procedure means the following. First we consider the
``negative'' plaquettes (the plaquettes with negative Tr $U_{\rm
plaq}$). The collection of such plaquettes represents a surface with a
boundary.  We add to this surface an additional surface in such a way
that the union of the two surfaces is closed.  In our procedure we
choose the additional surface so that it has minimal area. In other
words, we close the surface constructed from the ``negative''
plaquettes in a minimal way.  The resulting surface is the worldsheet
of the Simple Center Vortex.

It is obvious that the ``negative'' plaquettes in the continuum limit
become the singular configurations of the gauge field. Thus the Simple
Center Vortex is also a singular configuration.

Following \cite{su2} we construct the center monopoles (these objects are  known in the condensed matter physics as nexuses)
\begin{equation}
 j = \frac{1}{2} \; {}^*d [d N]\; {\rm mod}\, 2 .
\label{eq.07}
\end{equation}

\section{Scaling and asymptotic scaling in $SU(2)$ theory.}

The continuum limit of a lattice theory is obtained when we approach
the point of the second order phase transition. For the theory under
consideration this point is $\beta = \infty$, which means that the
correlation length $r(\beta)$ tends to infinity for $\beta \to
\infty$.  The physical correlation length remains of course the same,
but it becomes infinite in lattice units. In this situation any
physical object of finite length is represented on the lattice by an
infinite number of links. For example, let us write the physical
correlation length in lattice units $R_{\rm phys} = r(\beta) a(\beta)$,
where $a(\beta)$ is the physical size of a lattice link. In order to
keep the physical correlation length finite, the lattice spacing scales
as $a(\beta) \sim 1/ r(\beta)$ and consequently $a$ must tend to $0$
when we approach the phase transition and our lattice theory approaches
the continuum limit. This means that the lattice size scales as $L \sim
1/a$, when we keep the physical size of the  given lattice to be
independent of $\beta$.  The dependence of the lattice spacing $a$ on
$\beta$ is called scaling.  Suppose that some physical quantity which
is represented by some lattice variable $F_{\rm lat}$
 has the dimension $D$ in the units of mass, then $F_{\rm lat}
a^{-D} \rightarrow F_{\rm cont}$, where $ F_{\rm cont}$ is
this variable in the continuum limit. Thus we have for sufficiently large
$\beta$:
\begin{equation}
F_{\rm lat} \sim a^D
\end{equation}
A well-known example of such a behavior is the behavior of the string
tension: $\sigma_{\rm lat} a^{-2} \rightarrow \sigma_{\rm
cont}$.

The renormalization group analysis of the continuum theory predicts (up
to two - loop approximation) the following dependence of the lattice
spacing on $\beta$ \cite{Creutz}
\begin{equation}
 \bar{a}(\beta) \sim \beta^{\frac{51}{121}}e^{-(3\pi^2/11)\beta}
 \label{a}
\end{equation}
This behaviour is known as asymptotic scaling.

Thus we would like to see that for sufficiently large $\beta$ the
scaling of the lattice spacing approaches the asymptotic scaling. In
practice the asymptotic scaling in $SU(2)$ theory is not achieved (at least for
the values of $\beta$ from $2.1$ to $2.7$ which we used in the present
work). The deviations of the scaling from the asymptotic scaling for
these $\beta$ are well-known. One can extract the dependence of $a$
on $\beta$ from the lattice string tension. In the Tab.~\ref{tab.01} we represent the data
from Ref.~\cite{Scaling}.

\begin{table}
\caption{Behaviour of the string tension as a function of $\beta$:
 $\sqrt{\sigma_{\rm lat}} \sim \sqrt{\sigma_{\rm cont}}a(\beta)$.}
\label{tab.01} 
\begin{center}
\begin{tabular}{|c|c|c|c|}
\hline
  ~$N_\sigma$~ &~$N_\tau$~ &~$\beta$~ &~$\sqrt{\sigma_{\rm lat}}$~\\
\hline
 ~~8~&~10~&~2.20~&~0.4690(100)~\\
 ~10~&~10~&~2.30~&~0.3690(~30)~\\
 ~16~&~16~&~2.40~&~0.2660(~20)~\\
 ~32~&~32~&~2.50~&~0.1905(~~8)~\\
 ~20~&~20~&~2.60~&~0.1360(~40)~\\
 ~32~&~32~&~2.70~&~0.1015(~10)~\\
 ~48~&~56~&~2.85~&~0.0630(~30)~\\
\hline
\end{tabular}
\end{center}
\end{table}
 
One can check that $a(\beta)$ extracted from this data deviates from
$\bar{a}(\beta)$ for the values of $\beta$ considered.  It should be
mentioned that $\sqrt{\sigma_{\rm cont}}a$ is independent of the lattice size for
sufficiently large lattices. The data in Tab.~\ref{tab.01} is presented
for lattices of sizes $N_{\sigma}^3 N_{\tau}$.

\section{Fractal objects in the continuum limit of a lattice theory.}

In this section we consider the definition of a fractal object (See
also Refs.~\cite{fractal1, fractal}). We shall see that it follows from our
considerations in a straightforward way that objects of fractal
dimension $D>0$, as defined below, survive in the continuum limit.

If a one - dimensional object survives the continuum limit it must have
a length. We can introduce the following characteristic of this object:
The mean length of an object embedded into the unit four-volume, which
we denote by $\bar{l}$. The lattice density of these objects we denote
by $\rho$. Then
\begin{equation}
 \rho = N/L^4,
\end{equation}
where $N$ is the total number of elementary four-cubes covering 
our object inside a four-dimensional cube of lattice size
$L$.  The physical unit volume contains $L^4 \sim 1/a^4$ points of the
lattice.  The length of a linear object consisting of $N$ points is $N
a$, so the length of an object embedded into a four-dimensional cube of
lattice size L is $\rho L^4 a$.  Thus the length of a physical object
scales as $\bar{l} \sim \rho a^{1-4}$. It is important for us that
$\bar{l}$  is a real physical characteristic of a continuum object and
thus it should be independent of $\beta$ in the limit $\beta
\rightarrow \infty$.  That means that the lattice density of linear
object satisfies the equation
\begin{equation}
 \rho \sim a^{4-1}  .
\end{equation}
In the same way we obtain for a 2 - dimensional object surviving the
continuum limit:
\begin{equation}
 \rho \sim a^{4-2},
\end{equation} 
where $\rho$ is again the lattice density of these objects. 

For any integer $D$ we get for the $D$ - dimensional object:
\begin{equation}
 \rho \sim a^{4-D}  .
\label{D}
\end{equation} 
So, lattice objects with a lattice density that satisfies Eq.~(\ref{D})
at $\beta \rightarrow \infty$ can be considered as surviving the
continuum limit and having the dimension $D$.

When our object satisfies Eq.~(\ref{D}) with noninteger $D>0$, we treat
it as a fractal object of dimension $D$. This point of view becomes
transparent after the demonstration that the above definition of the
fractal dimension is in accordance with the definition of the Hausdorff
dimension of a set embedded in four-dimensional space.

The Hausdorff dimension of an object in the four-dimensional continuum
is defined in the following way \cite{fractal1}: Consider a
four-dimensional cube of fixed physical size. Subdivide this cube into
$L^4$ different subcubes.  The number of subcubes covering our object
is denoted by $N$. If $N \sim L^D$ and $D>0$ at $L \rightarrow \infty$
we say that our object has Hausdorff dimension $D$.

How can we represent the subdivision of a cube of some physical size
into different numbers of subcubes using the lattice theory?  The
answer is as follows. The subdivision into the infinite number of
subcubes is represented via the continuum limit itself (the lattice
theory at $\beta = \infty$). The lattice theory for finite $\beta$ is
not equal to the continuum theory. But it becomes closer to the
continuum limit when $\beta$ becomes larger. Instead of the subdivision
of the cube into $L^4$ subcubes in the continuum theory we can use the
lattice theory defined on the lattice of size $L$. We have already seen
that the size of the lattice which represents the same physical volume
scales as $L \sim 1/a(\beta)$, where $a(\beta)$ represents the scaled lattice spacing. Thus the fractal dimension of some object (up to the difference between the
pure continuum theory and it's lattice version for large $\beta$) can
be extracted from the formula
\begin{equation}
N \sim L^{D}, \label{LD}
\end{equation}
where $N$ is the number of cubes, which cover our object inside the
lattice of size $L$. The difference between the two theories disappears
at $L \rightarrow \infty$ (which implies that $\beta \rightarrow
\infty$).  Thus if Eq.~(\ref{LD}) is valid for $\beta \rightarrow \infty$,
when $L$ and $N$ are treated as  functions of $\beta$  while $D$
remains independent of $\beta$, we can consider $D$ to be the fractal
dimension of our object existing in the continuum theory.

Now let us show that an object on the lattice, which density scales for
$\beta \to \infty$ as in Eq.~(\ref{D}) with $D>0$ can be considered as
an object in the continuum with Hausdorff dimension $D$.  The number
$N$ of subcubes covering the elements of our object is involved into
the definition of the lattice density: $\rho = N/L^4$.  Thus the number
of elementary subcubes covering our object inside the four-dimensional
cube of some fixed physical size can be written as 
\begin{equation}
 N \sim \rho L^4, 
\end{equation} 
where $L \sim 1/a(\beta)$. We get
from Eq.~(\ref{D}):  
\begin{equation} \rho = N/L^4 \sim a^{4-D}.
\end{equation} 
Thus 
\begin{equation}
 N \sim L^{D}.  
\end{equation} 
Here $D$ is independent of $\beta$.
According to above we can treat it as the Hausdorff dimension.

We summarize this section as follows: If an object under
consideration has a lattice density which satisfies Eq.~(\ref{D}) for
$\beta \rightarrow \infty$, we say that this object survives the
continuum limit and can be treated as a fractal object of dimension
$D$. The definition of such a dimension is in accordance with the
definition of Hausdorff dimension.

\section{The reality of the existence of the vortices in the continuum
limit.}

In this section we represent our numerical results. We made our
simulations using lattices of sizes $16^4$ and $24^4$.  We found no
difference between the results obtained on those lattices, which is our
reason to believe that the lattice size has no influence on the
considered quantities at all, for lattices of size $16^4$ and greater.

\subsection{The density of the vortices and the monopoles.}

\begin{figure}
\begin{center}
 \epsfig{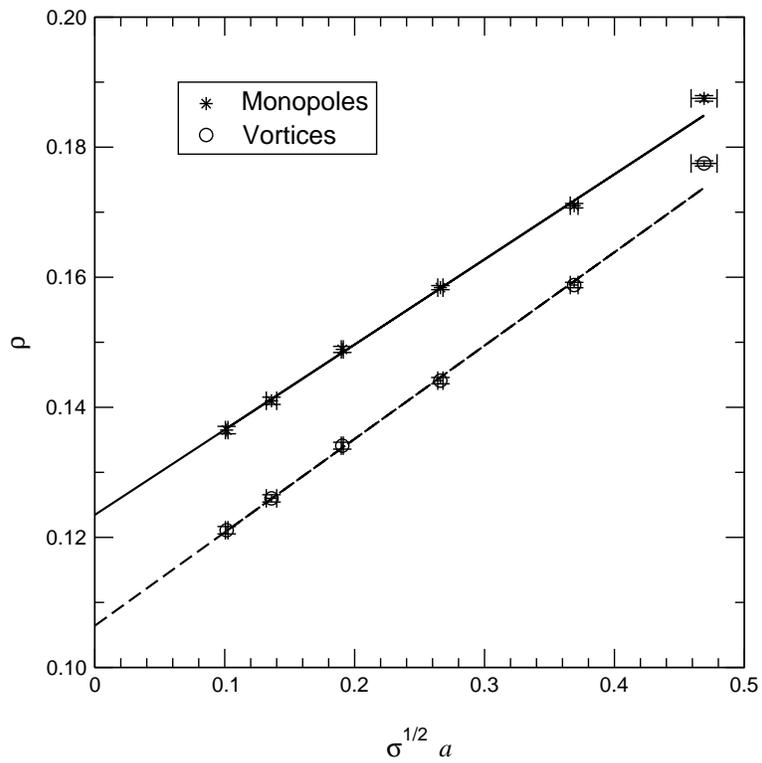}

\vspace{1ex}

 \caption{\label{fig.01} The dependence of $\rho$ on $a(\beta)$ for the
 Simple Center Vortices and the Simple Center Monopoles.
 The lattice has dimensions $24^4$. The lines are linear fits.
 \label{fig.10}}
\end{center}
\end{figure}

The numerical investigation of the lattice density of the Simple Center
Vortices and of the center monopoles are represented in
Fig.~\ref{fig.10}. We represent $\rho$ as a function of $a(\beta)$. The values of $\sqrt{\sigma_{\rm cont}} a(\beta)$ for finite particular $\beta$ are represented in the Tab.~\ref{tab.01}.  From
Fig.~\ref{fig.10} we find that within the errors the dependence is indeed
linear
\begin{equation}
\rho = \rho_c + \alpha\, a(\beta) .
\end{equation}
Here $\rho_c$ is the density at $a(\beta) = 0$ obtained via the
extrapolation of the data from Fig.~\ref{fig.10}. For the center monopoles we
we find: $\rho_c = 0.123\pm 0.001$ and for the vortices  $\rho_c =0.106\pm
0.001 $. It is clear now that $\rho (\beta)$ does not tend to $0$ for
$\beta \to \infty$.
Thus we have, both for the vortices and the monopoles.
\begin{equation}
 \rho \sim a^{4-D} 
\end{equation} 
with $D = 4$. So we find for our objects a specific behavior of the
density. It does not vanish for $\beta \to \infty$.  That means
according to the previous section, that those objects survive the
continuum limit with a fractal dimension equal to $4$.

It should be mentioned here that the action near the monopole current is
greater than the average value of the action calculated over all the
lattice. The excess varies between 5\% and 8\% in the interval $2.1 <
\beta < 2.6$. It means that this object carries energy.  The monopoles
form one big cluster and several small ones. This situation is similar
to the maximal Abelian case\cite{BCGPSVZ}. Following this reference we
call the large loops infrared and the small ones ultraviolet monopoles.
The latter are unphysical. We found that the fraction of unphysical
monopoles amounts to about $1$ - $3$\% of all the monopoles.

\subsection{The Creutz ratio without simple center vortices. }
\begin{figure}
\begin{center}
 \epsfig{figure=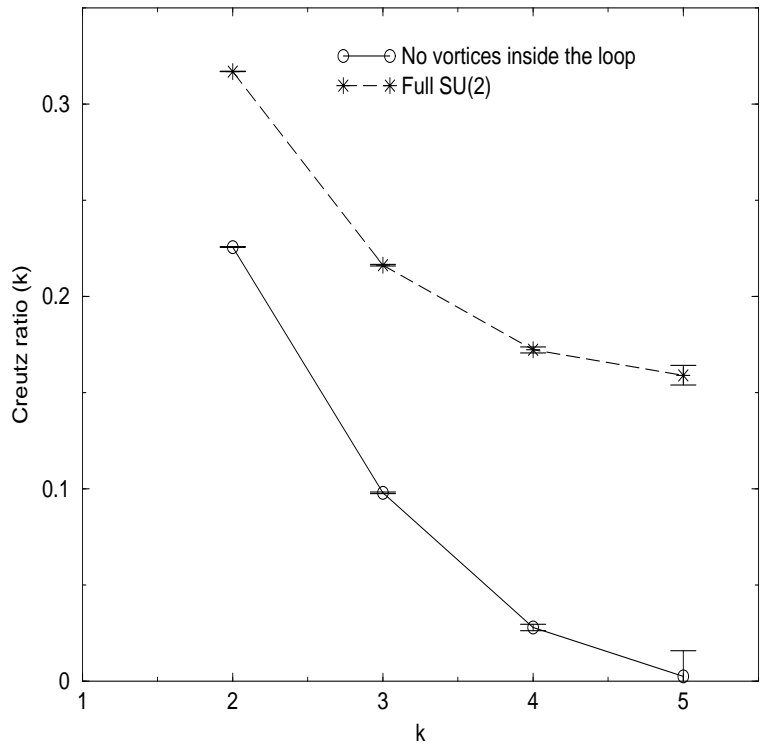,height=100mm,width=100mm,angle=-90}
 \caption{\label{fig.02} The Creutz ratio for the configurations for
 which there is no vortex inside the Wilson loops versus the size of
 the loop. The Creutz ratio for full SU(2) is given for comparison.
 \label{fig.20}}
\end{center}
\end{figure}

To illustrate that the string tension is due to the Simple Center
Vortex we consider the configurations for which there are no vortices
inside the Wilson loop considered. In Fig.~\ref{fig.20} the
dependence of the Creutz ratio for such configurations on the size of the
Wilson loop is represented for $\beta =2.3$.  This Creutz ratio
vanishes for large size $k$. We have found that the same result takes
place for all values of $\beta$. On the other hand, the string tension
does not vanish in the continuum limit. That means that the singular
Simple Center Vortices play a real role in the dynamics.

\section{Conclusions}
In this work we are trying to answer the question: ``Do singular
configurations live in the continuum $SU(2)$ theory and do they play a
real role in the dynamics?'' Our answer is ``Yes.'' That means that
these singular field configurations should be taken seriously in the
investigation of gauge-field theory.

\vspace{10mm}

We are grateful to J.Greensite, M.Faber, and M.I. Polikarpov for useful
discussions. A.I.V. and M.A.Z. kindly acknowledge the hospitality of the
Department of Physics and Astronomy of the Vrije Universiteit, where
part of this work was done.  This work was partly supported by RFBR
grants 02-02-17308 and 01-02-17456, INTAS 00-00111 and
CRDF award RP1-2364-MO-02.

\end{document}